# Frequency analysis and resonant operation for efficient capacitive deionization

Ashwin Ramachandran,[a] Steven A. Hawks,[c] Michael Stadermann,[c] Juan G. Santiago [b,*]

[a] Department of Aeronautics & Astronautics, Stanford University, Stanford, California 94305, United States

[b] Department of Mechanical Engineering, Stanford University, Stanford, California 94305, United States

[c] Lawrence Livermore National Laboratory, 7000 East Avenue, Livermore, California 94550, United States

* To whom correspondence should be addressed. Tel. 650-736-1283, Fax 650-723-7657, E-mail: juan.santiago@stanford.edu

## Abstract

Capacitive deionization (CDI) performance metrics can vary widely with operating methods. Conventional CDI operating methods such as constant current and constant voltage show advantages in either energy or salt removal performance, but not both. We here develop a theory around and experimentally demonstrate a new operation for CDI that uses sinusoidal forcing voltage (or sinusoidal current). We use a dynamic system modeling approach, and quantify the frequency response (amplitude and phase) of CDI effluent concentration. Using a wide range of operating conditions, we demonstrate that CDI can be modeled as a linear time invariant system. We validate this model with experiments, and show that a sinusoid voltage operation can simultaneously achieve high salt removal and strong energy performance, thus very likely making it superior to other conventional operating methods. Based on the underlying coupled phenomena of electrical charge (and ionic) transfer with bulk advection in CDI, we derive and validate experimentally the concept of using sinusoidal voltage forcing functions to achieve resonance-type operation for CDI. Despite the complexities of the system, we find a simple relation for the resonant time scale: the resonant time period (frequency) is proportional (inversely proportional) to the geometric mean of the flow residence time and the electrical (*RC*) charging time. Operation at resonance implies the optimal balance between absolute amount of salt removed (in moles) and dilution (depending on the feed volume processed), thus resulting in the maximum average concentration reduction for the desalinated water. We further develop our model to generalize the resonant time-scale operation, and provide responses for square and triangular voltage waveforms as two examples. To this end, we develop a general tool that uses Fourier analysis to construct CDI effluent dynamics for arbitrary input waveforms. Using this tool, we show that most of the salt removal (~95%) for square and triangular voltage forcing waveforms is achieved by the fundamental Fourier (sinusoidal) mode. The frequency of higher Fourier modes precludes high flow efficiency for these modes, so these modes consume additional energy for minimal additional salt removed. This deficiency of higher frequency modes further highlights the advantage of DC-offset sinusoidal forcing for CDI operation.

## 1. Introduction

Desalination using capacitive deionization (CDI) is an emerging and attractive technology for brackish water treatment (Anderson et al., 2010; Oren, 2008; Suss et al., 2015; Welgemoed and Schutte, 2005). Like many other multi-physics problems, CDI involves coupling of multiple time scales and phenomena (Johnson and Newman, 1971). CDI salt removal dynamics are determined by the interplay between electrical charging/discharging (which depends on cell ionic and electrical resistances and capacitance) coupled with bulk mass transport (Biesheuvel et al., 2009; Guyes et al., 2017; Hemmatifar et al., 2015; Qu et al., 2018). Moreover, CDI is inherently periodic because electrical charging and discharging forcing functions result in periodic salt removal and regeneration phases.

CDI performance can be evaluated using a recently proposed set of metrics (Hawks et al., 2018b). These performance metrics include average concentration reduction, volumetric energy consumption, and productivity for 50% water recovery. Owing to the multiphysics nature of CDI, the desalination performance can be affected dramatically by the particular choice of operating method. Most of the previous research on CDI operation has centered around the use of constant current (CC) and/or constant voltage (CV), and very little attention has been given to other possible operational schemes. CC operation has been shown to consume less energy compared to CV, given equal amount of salt removal (Choi, 2015; Kang et al., 2014; Qu et al., 2016). CC can also achieve a controllable quasi-steady state effluent concentration (Hawks et al., 2018a; Jande and Kim, 2013). Conversely, CV can achieve faster rates of desalination, albeit with a tradeoff in energy consumption (Wang and Lin, 2018a, 2018b). Recent research around operational schemes for CDI have proposed mixed CC-CV modes (García-Quismondo et al., 2013; Saleem et al., 2016), variable flow rate (Hawks et al., 2018a), changing feed concentration (García-Quismondo et al., 2016), and variable forcing function periods (Mutha et al., 2018). Generally, these studies can be characterized as ad-hoc operational strategies geared toward the improvement of one (or few) metrics at the cost of others.

We know of no work which combines a theoretical framework and accompanying validation experiments which explore generalized control waveform shapes for CDI. In other words, to date, studies have only explored ad hoc operational schemes such as square waves in applied current or voltage. A key step in developing good operation modes for CDI would involve understanding the role of arbitrary periodic forcing functions (including frequency and wave shape) on the aforementioned desalination performance metrics.

In this work, we show that CDI desalination dynamics can be, under appropriate operation conditions, modeled as a linear time invariant system. Further, we propose, describe, and demonstrate a new operation scheme for CDI that uses either a sinusoidal forcing voltage (our preferred method in this study) or sinusoidal current. In particular, we highlight several advantages of using a sinusoidal forcing for CDI as compared to conventional operation methods. To our knowledge, our work is the first to introduce and quantify the performance of a sinusoidal forcing function for CDI. This sinusoidal forcing results in an approximately sinusoidal effluent concentration with an amplitude, phase, and waveform that can be predicted accurately. We use theory and experiments to show sinusoidal forcing can be modeled with a dynamic systems approach and that there exists a system-inherent and "resonant" time scale that strongly enhances the desalination performance of CDI, while simultaneously achieving good energy performance. As an example, we analyze and compare this sinusoidal forcing to more traditional constant voltage (square wave) and triangular voltage waveforms. Further, in Appendix A1, we present example engineering design approaches and associated expected performance metrics for CDI operation at resonance. Finally, we present a generalized framework that uses Fourier analysis to construct responses for CDI for arbitrary input current/voltage forcing functions. The tools presented here can be applied to analyzing a wide range of CDI operations, quantifying performance, and CDI system optimization.

## 2. Theory – A resonant CDI operation

We here formulate a theory around CDI desalination dynamics for a sinusoidal forcing current or voltage. For simplicity and without significant loss of applicability, we treat the electrical response of the CDI cell as a simple, series, linear RC circuit with effective $R$ and $C$ values, as determined in Section 3.2 (see Hawks et al., 2018a; Ramachandran et al., 2018 for details). The electrical forcing of the CDI cell results in a desalination response in terms of an effluent concentration versus time. Again for simplicity, we describe the coupling between electrical input and concentration of the output stream using a simple continuous stirred-tank reactor (CSTR) model (Biesheuvel et al., 2009; Hawks et al., 2018a; Jande and Kim, 2013; Ramachandran et al., 2018).

For the CDI cell electrical circuit, we assume a DC-offset sinusoidal forcing voltage given by

$$V(t) = V_{dc} + \Delta V \sin(\omega t), \tag{1}$$

where $V_{dc}$ is the constant DC component of applied voltage (typically $> 0$ V for good performance; see Kim et al., 2015 for a related discussion), $\Delta V$ is the amplitude of the sinusoid voltage, and $\omega$ is the forcing frequency. Under dynamic steady state (DSS) such that the initial condition has sufficiently decayed as per the CDI system's natural response (Ramachandran et al., 2018), current $I$ in the electrical circuit is obtained as (see SI Section S1 for derivation),

$$I(t) = \frac{C\Delta V \omega}{\sqrt{1 + (\omega RC)^2}} \cos(\omega t - \arctan(\omega RC)) = \frac{C\Delta V \omega}{\sqrt{1 + (\omega RC)^2}} \sin\left(\omega t + \frac{\pi}{2} - \arctan(\omega RC)\right). \tag{2}$$

We can represent the result in equation (2) as

$$I(t) = \Delta I \sin(\omega t + \phi_{IV}) \tag{3}$$

where the amplitude, and the phase of current with respect to voltage are given by

$\Delta I = \dfrac{C\Delta V \omega}{\sqrt{1 + (\omega RC)^2}}$, and $\phi_{IV} = \dfrac{\pi}{2} - \arctan(\omega RC)$, respectively.

Further, we describe the dynamics that govern effluent concentration reduction $\Delta c$ via the mixed reactor model approximation (Ramachandran et al., 2018) as,

$$\tau \frac{d(\Delta c)}{dt} + \Delta c = \frac{I(t)\overline{\Lambda}}{FQ} \qquad (4)$$

where $\Delta c = c_0 - c(t)$ represents an appropriate reduction of the feed concentration $c_0$ at the effluent, $Q$ is flow rate, $F$ is Faraday's constant, $\tau \ (= \forall / Q)$ is the flow residence time scale ($\forall$ is the mixed reactor volume), and $\overline{\Lambda} \ (= \lambda_{dl}\lambda_c)$ is an effective dynamic charge efficiency ($\lambda_c$ and $\lambda_{dl}$ are respectively the cycle averaged Coulombic and EDL charge efficiencies); see Hawks et al., 2018a and Ramachandran et al., 2018 for further details. Using equation (3) in (4), and solving for effluent concentration reduction under DSS, we obtain

$$\Delta c(t) = \frac{C \Delta V \omega \overline{\Lambda}}{FQ \sqrt{1 + (\omega \tau)^2} \sqrt{1 + (\omega RC)^2}} \sin \left( \omega t + \frac{\pi}{2} - \arctan \left( \frac{\omega (RC + \tau)}{1 - \omega^2 \tau RC} \right) \right). \qquad (5)$$

Equivalently,

$$\Delta c(t) = \Delta c_{ac} \sin \left( \omega t + \phi_{cV} \right) \qquad (6)$$

where $\Delta c_{ac} = \dfrac{C \Delta V \overline{\Lambda} \omega}{FQ \sqrt{1 + (\omega \tau)^2} \sqrt{1 + (\omega RC)^2}}$ is the maximum change in effluent concentration, and

$\phi_{cV} = \dfrac{\pi}{2} - \arctan \left( \dfrac{\omega (RC + \tau)}{1 - \omega^2 \tau RC} \right)$ is the phase of $\Delta c$ with respect to $V$. Further, the phase of $\Delta c$ with respect to current $I$ is given by $\phi_{cI} = -\arctan(\omega \tau)$. Also, the average concentration reduction at the effluent is given by $\Delta c_{avg} = 2 \Delta c_{ac} / \pi$, and water recovery is 50%. Note the absolute concentration difference $\Delta c_{ac}$ depends on extensive (versus mass-specific, intensive) CDI cell properties such as *R*, *C,* and cell volume.  Importantly, $\Delta c_{ac}$ is also a function of operational parameters such as *Q*, voltage window, and forcing frequency $\omega$. We find that the basic coupling of RC circuit dynamics and mixed reactor flow directly results in what we here will refer to as a "resonant frequency", $\omega_{res}$.  This frequency maximizes effluent concentration reduction $\Delta c_{ac}$ in Equation (6) and is simply the inverse geometric mean of the respective circuit and flow time scales,

$$\omega_{res} = \frac{1}{\sqrt{\tau RC}} \ .$$

(7)

Furthermore, the maximum average concentration reduction $\Delta c_{avg,res}$ achieved at the resonant frequency is given by,

$$\Delta c_{avg,res} = \Delta c_{avg}\Big|_{\omega=\omega_{res}} = \frac{2C\Delta V \overline{\Lambda}}{\pi F \forall} \frac{1}{1+\left(RC/\tau\right)} = \frac{\overline{\Lambda}}{\pi F \forall} \frac{\Delta V}{R}\left(\frac{2\tau RC}{\tau+RC}\right) \ .$$

(8)

For a given cell (fixed *R, C,* and $\forall$), Eq. (8) is an expression which can be used to design a sinusoidal operation (i.e., appropriate choice of flowrate and voltage window) to achieve a certain desalination depth $\Delta c_{avg}$. Refer to Appendix A1 for a discussion on this design approach and the energy and throughput metrics associated with operation at resonance. Our rationale behind the term "resonant/resonance" is explained as follows. CDI as a periodic dynamic process involves the coupling of several physical phenomena including (i) electrical charging/discharging (governed by the *RC* time scale), with (ii) salt removal at the electrodes and freshwater recovered at the outlet by fluid flow (governed by the flow residence time scale $\tau$); see Figure 1. Each of these two time scales affects salt removal, and is physically independent of the other.  The CDI system's average concentration reduction $\Delta c_{avg} = \langle \Delta c(t) \rangle$ (where the brackets indicate a time average over the desalination phase) therefore couples the two time scales in a manner very similar to resonance in a dynamic system. Hence, we refer to the periodic CDI operation at the fundamental frequency $\omega_{res}$ (independent of the forcing function waveform) as a "resonant operation".

Lastly, we note that the dynamic system analysis presented in this section can also be derived using a Laplace transform formulation involving transfer functions for the CDI system. For readers who may find this more intuitive or familiar, we provide such a formulation in Section S1 of the SI. Somewhat surprisingly, the present work is the first to develop such transfer function formulation for practical operations using CDI.

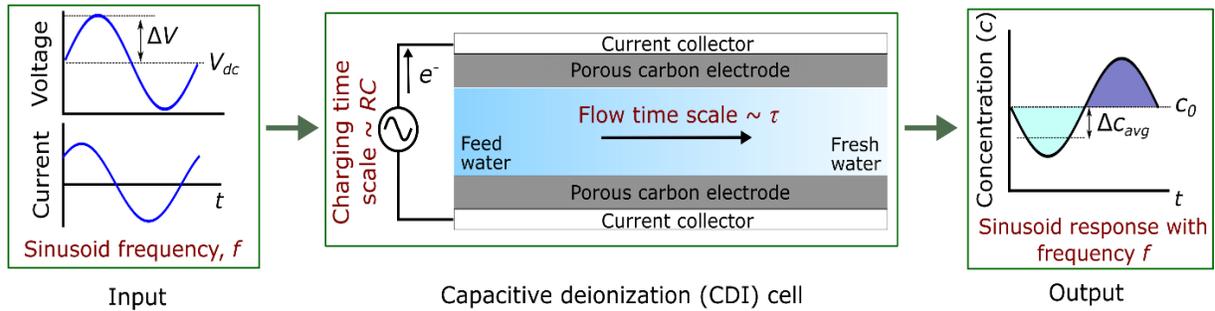

**Figure 1.** Schematic of CDI operation and the physical time scales involved in determining desalination performance. The figure has three parts: the *input* represents a sinusoidal forcing voltage (current) applied to the CDI *system*, which results in a sinusoidal time variation of effluent concentration as the *output*. A second output is system current (voltage). Note that the two essential time-scales given by the electronic time *RC* and flow residence time $\tau$ together determine output response and hence CDI performance including degree of desalination, power consumed, and productivity.

## 3. Materials and Methods

### 3.1 CDI cell design

We fabricated and assembled a flow between (fbCDI) cell using the radial-flow architecture described in Biesheuvel and van der Wal, 2010, Hemmatifar et al., 2016, and Ramachandran et al., 2018. Five pairs of activated carbon electrodes (Materials & Methods, PACMM 203, Irvine, CA) with 5 cm diameter, 300 µm thickness, and total dry mass of 2.65 g were stacked between 5 cm diameter, 130 µm thick titanium sheets which acted as current collectors. We used two 180 µm thick non-conductive polypropylene circular meshes (McMaster-Carr, Los Angeles, CA) between each electrode pair as spacers, with an estimated porosity of ~59%. The spacers had a slightly larger (~5 mm) diameter than the electrodes and current collectors to prevent electrical short circuits.

### 3.2 Experimental methods and extraction of model parameters

The experimental setup consisted of the fbCDI cell, a 3 L reservoir filled with 20 mM potassium chloride (KCl) solution which was circulated in a closed loop, a peristaltic pump (Watson Marlow 120U/DV, Falmouth, Cornwall, UK), a flow-through conductivity sensor (eDAQ, Denistone East, Australia) close to the cell outlet, and a sourcemeter (Keithley 2400, Cleveland, OH). We estimate less than 1% change in reservoir concentration based on adsorption capacity of our cell, and thus approximate influent concentration as constant.

The resistance and capacitance of the cell were characterized using simple galvanostatic charging, and these estimates were corroborated by electrochemical impedance spectroscopy (EIS) and cyclic voltammetry measurements using a potentiostat/galvanostat (Gamry Instruments, Warminster, PA, USA)); see SI Section S2 for data. We estimated a differential cell capacitance of $33\pm1.8$ F (equivalently $\sim$ 44 F/cm$^3$ and 49 F/g) and an effective series resistance of $2.85\pm0.28$ Ohms, resulting in a system $RC$ time scale of ~94 s. To determine the mixed reactor cell volume $\forall$, we used an exponential fit to the temporal response (open-circuit flush) of the cell as described in Hawks et al., 2018a and Ramachandran et al., 2018, and we estimated $\forall$ of 2.1±0.2 ml. For simplicity, all of the forced (sinusoidal, triangular and square voltage) responses presented in this work are at a constant flowrate of 2.3 ml/min, corresponding to a residence time scale $\tau$ (=$\forall / Q$) of ~55 sec. Thus, the operational and system parameters described here result in a resonant frequency $f_{res}$ (=$\omega_{res} / 2\pi$) value of 2.2 mHz (using Eq. (7)), and a corresponding resonant time scale $T_{res}$ ($= 1/ f_{res}$) of 450 sec. The water recovery was 51-57% for all the cases presented here.

**Section 4. Results and Discussion**

*4.1 CDI as a first order linear time invariant (LTI) dynamic process – response to sinusoid voltage forcing*

We here study the desalination dynamics associated with CDI from a "dynamical system modeling" viewpoint. To this end, we subject the CDI cell with a constant flow rate and operate with a sinusoidal voltage forcing. Further, we constrain the voltage of operation within reasonable limits: sufficiently low peak voltage such that the Coulombic losses are small, and a voltage window such that EDL charge efficiency can be approximated by a constant value (Hawks et al., 2018a; Kim et al., 2015; Ramachandran et al., 2018).

Figure 2 shows a plot of experimental data along with a corresponding prediction by the model (c.f. Section 2). Plotted is the effluent concentration $c$ versus time for a sinusoidal voltage operation with a voltage window of 0.7 to 1.1 V, and a constant flowrate of 2.3 ml/min. Results are shown for three different frequencies approximately spanning a decade (0.9, 2.5, and 8.8 mHz). For experimental data, a time delay of ~ 4 s was subtracted from the measured time, which is associated with the temporal delay associated with transport and dispersion between cell concentration and the downstream conductivity meter. For the model, we used a constant value of EDL charge efficiency of 0.91 (determined using data shown for the same voltage window in Figure 3), and used an experimentally determined (average) value of Coulombic efficiency of 0.94 (a value we found to be nearly constant for the all frequencies shown in Fig. 2). Using the sinusoidal voltage forcing (shown in inset of Fig. 2), we observed that the measured effluent concentration also varies, to very good approximation, as a sinusoidal in time. Further, our model predicts experimental observations (both amplitude and phase of $c$) very well over the range of frequencies presented in Fig. 2. The observation that a sinusoidal forcing function (here, voltage or current) to a dynamical system (here, the CDI cell) results in a nearly sinusoidal response (here, the effluent concentration) is, of course, a characteristic of an approximately linear time invariant (LTI) system. By definition, an LTI system is both linear and time-invariant, i.e., the output is linearly related to the input, and the output for a particular input does not change depending on when the input was applied.

We thus infer that the desalination dynamics using CDI can be modeled to a good approximation as a first order linear time invariant (LTI) system under the following conditions: (i) constant flowrate (with advection dominated transport), (ii) small variation in dynamic EDL charge efficiency such that it can be approximated by a constant value, and (iii) high Coulombic efficiency (close to unity). LTI systems have well-developed tools for system analysis and control (Franklin et al., 2002), and thus can be applied to analyzing CDI systems. In Section S3 of the SI, we provide one anecdotal "off-design" sinusoidal input operation of CDI which results in significant distortion of the output concentration. Namely, we show the case of a large variation in EDL charge efficiency due to a large voltage window wherein effluent concentration exhibits a significant deviation from a sine wave. We hope to further study such deviations from linearity in future work.

Importantly, the predictions and experimental data of Figure 2 show that the effluent concentration has a frequency-dependent amplitude and a distinct phase shift with respect to the forcing voltage waveform—an observation which we study further in Section 4.2.

Lastly, we note that, although we here focus on sinusoidal voltage forcing functions, our work with the present model suggests sinusoidal current can also be used to characterize CDI dynamics. We hypothesize that sinusoidal applied current can also yield sinusoidal time variation of effluent concentration, thus extending the present work. We performed some preliminary experiments toward such a study and observed that sinusoidal forcing currents easily lead to deviations from ideal behavior (and the model) due to unwanted Faradaic (parasitic) reactions. This results in an attenuation of concentration reduction in regions of high voltage, and a more complex natural response relaxation from the initial condition. Such sinusoidal forcing also requires non-zero DC values for applied current to account for unavoidable Faradaic losses. We thus prefer a sinusoidal voltage over sinusoidal current forcing as a more controllable and practical operating method.

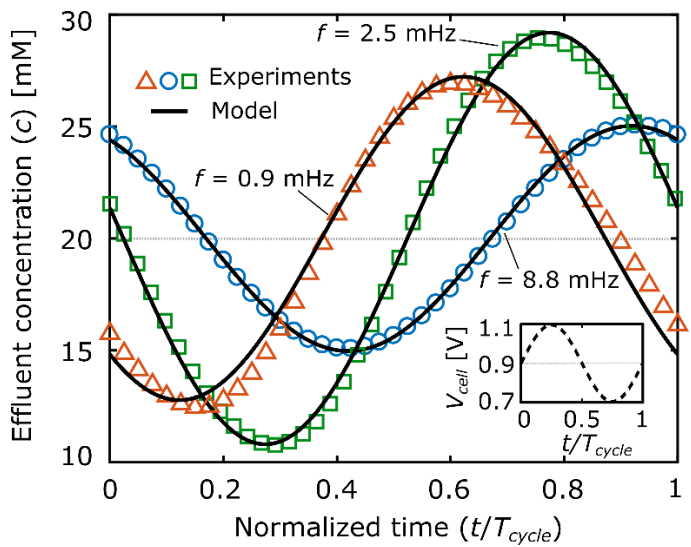

**Figure 2.** Effluent concentration versus time (normalized by cycle period) of the CDI cell for a sinusoidal voltage input between 0.7 - 1.1 V ($V_{dc} = 0.9$ V and $\Delta V = 0.2$ V) with frequencies of 0.9, 2.5 and 8.8 mHz, a constant flowrate of 2.3 ml/min, and a feed concentration of 20 mM. Symbols and solid lines respectively correspond to experimental data and model results. Inset shows the sinusoidal voltage forcing function. Note that under dynamic steady state (DSS), the time variation of effluent concentration of the CDI cell is approximately sinusoidal (over a wide range of frequencies) for a sinusoidal voltage forcing function. Thus, under appropriate operating conditions, CDI can be modeled as a linear time invariant (LTI) dynamical process.

### 4.2 Frequency response: Bode plot and resonant frequency analysis for CDI

In this section, we present a frequency analysis of the response of current and effluent concentration in CDI for a forcing sinusoidal voltage. Figures 3a and 3b show measured current and effluent concentration profiles versus time (normalized by cycle duration) for a sinusoidal voltage forcing with frequencies spanning 0.2 to 17.7 mHz. Shown in Figs. 3a and 3b are results for two voltage windows (see inset of Fig. 3a) with the same $\Delta V$ of 0.2 V, but with $V_{dc}$ values of 0.8 V (dashed lines) and 0.9 V (solid lines). Figures 3c and 3e respectively show the frequency dependence of the amplitude and phase of the current response (i.e. Bode plots for current). Figures 3d and 3f show the corresponding frequency dependence of average concentration reduction and phase shift in effluent concentration (Bode plots for $\Delta c$). Note that for data in Figs. 3c and 3e, we choose the governing *RC* time scale (for current response) for normalizing the frequency, and for effluent concentration data in Figs. 3d and 3f we choose the resonant time scale (which governs $\Delta c$) to normalize frequency.

### 4.2.1 Current response

From Fig. 3a, we notice that the current response for a sinusoidal forcing voltage to the CDI cell is also sinusoidal (to a good approximation) over a wide range of frequencies. We quantify the amplitude and phase lead of the current response from experiments versus the forcing frequency (normalized by the $RC$ frequency) in Figs. 3c and 3e respectively. For calculating amplitude, we average the two peak values of current (corresponding to charging and discharging) after subtracting the DC value (corresponding to leakage current at $V_{dc}$). For calculating phase shift of current with respect to forcing voltage from data, we averaged the two phase shifts estimated using the time delay (normalized by cycle time) between the peak values of the sinusoidal current and voltage. We further overlay results from the model in Figs. 3c and 3e.

Notice in Fig. 3e that current always leads the forcing voltage in time (i.e., $\phi_{IV} > 0$), as expected for an RC-type electrical circuit. In other words, the peak in current response occurs before the corresponding peak value of forcing voltage. Further, the phase lead of current with respect to voltage decreases with increasing frequency (c.f. the shift in the sinusoidal current profile to the right in Fig. 3a). At $f = f_{RC} = (RC)^{-1}$, the phase lead of current is ~45 degrees. Note also from Figs. 3a, 3c and 3e that operationally, the current profile (amplitude and phase shift) is less sensitive to the DC voltage ($V_{dc}$) value, since it mainly depends on $\Delta V$, and system parameters $R$ and $C$ (from Eq. (3)). Also, note the good agreement of our model predictions for both amplitude and phase of current, especially for the most practically relevant, moderate-to-low frequency range of operation. We hypothesize that the deviation of our model predictions from experiments at high frequencies ($f \geq 3 f_{RC}$) is due to a deviation from a constant $RC$, linear assumption. At these relatively high frequencies, the CDI cell electrical response exhibits a transient response better modeled using more complex circuits such as the transmission line response associated with non-linear distributed EDL capacitances (de Levie, 1963; Qu et al., 2016; Suss et al., 2012).

### 4.2.2 Effluent concentration response

We here follow an averaging procedure similar to that of Section 4.2.1 to evaluate the phase and amplitude of the effluent response. For the effluent response, the only fitting parameter for the model is the product $\overline{\Lambda}$ ($=\lambda_{dl}\lambda_c$), and we determine this product from the aforementioned best fit curve approach to extract cycle-averaged Coulombic and double layer efficiencies from the experimental data (see SI Section S4 for further details). We obtained values of $\overline{\Lambda}$ of 0.8 (corresponding to $\lambda_{dl}$ of 0.91 and $\lambda_c$ of 0.88) and 0.73 (corresponding to $\lambda_{dl}$ of 0.82 and $\lambda_c$ of 0.92) for $V_{dc}$ of 0.9 V and 0.8 V, respectively. Unlike the monotonic variations of phase and amplitude observed for current response in Figs. 3c and 3e, effluent concentration exhibits a distinctly non-monotonic variation in amplitude with changing frequency. From Figs. 3b and 3d, we observe that as frequency increases, the amplitude of effluent concentration variation (and the average concentration reduction) increases, reaches a maximum, and then decreases. Further, unlike current, the effluent concentration profile both leads ($\phi_{cV} > 0$) and lags ($\phi_{cV} < 0$) the forcing voltage at low and high frequencies, respectively, as shown in Fig. 3f. The "special" frequency that corresponds to both (i) *maximum amplitude*, and (ii) the change in sign of the phase of effluent concentration with respect to the forcing voltage, is the resonant frequency $f_{res}$. At this resonant frequency, the effluent concentration is *exactly in phase* with the forcing sinusoid voltage function.

Operation at the resonant frequency results in the maximum desalination depth $\Delta c_{avg}$ for a given voltage window, which is clearly supported by experiments and model results shown in Fig. 3d.

Also, note that $\Delta c_{avg}$ drops by ~50% for a frequency that is a factor of 5 away from the resonant frequency. Unlike current, the effect of voltage $V_{dc}$ (for the same $\Delta V$) on the amplitude of $\Delta c$ is significant, as shown in Figs. 3b and 3d. Specifically, for the same $\Delta V$, a higher $V_{dc}$ (within the Faradaic dominant voltage limit of ~1.2 V, such that Coulombic efficiency is close to unity) results in a higher EDL efficiency (and thus cycle averaged charge efficiency). This yields higher $\Delta c_{avg}$ as per Eq. (5). Conversely, the phase shift in effluent concentration is relatively insensitive to $V_{dc}$ (Fig. 3f). As with the current response data, our effluent amplitude and phase measurements deviate from the model at higher frequencies $(f \geq 3 f_{res})$. We hypothesize that this is primarily due to the inaccuracy of the mixed flow reactor formulation (for cycle times significantly lower than the flow residence time).

### 4.2.3 Physical significance of the resonant frequency and operation: Limiting regimes

CDI as a practical dynamic process most often involves two dominant and independent time scales: (i) an $RC$ time (electronic time scale associated with electrical circuit properties), and (ii) flow residence time (ionic transport time scale in a mixed reactor volume). The interplay between these two time scales determines the desalination depth $\Delta c_{avg}$ at the effluent. To better understand this interplay, we here describe operating scenarios corresponding to very high and very low operating frequencies.

At high frequency operation $(f \gg f_{\tau} \textbf{ and } f_{RC})$, the rapid forcing results in repeated desalination and regeneration (salt uptake and rejection) from and to approximately the same volume of water contained in the CDI cell. Further, the RC-type electrical response of the cell is such that high frequencies incompletely charge the capacitive elements of the cell. This wasteful operation consumes energy and leads to low $\Delta c_{avg}$. For very low frequencies $(f \ll f_\tau \textbf{ and } f_{RC})$ or equivalently long cycle durations, the EDLs are fully charged (high EDL charge efficiency) and freshwater recovery at the effluent is high (flow efficiency close to unity; c.f. Section 4.3.2); each of which is favorable. However, in this limiting regime, the system can be characterized as suffering from the mitigating effect of "overly dilute" effluent. That is, after EDL charging, the majority of the charging phase is spent flushing feed water through (and out of) the cell. Similarly, after EDL discharge, the majority of the discharging phase is again spent flowing feed water. Both of these phases hence exhibit a low value of the inherently time-averaged magnitude of $\Delta c_{avg}$. Note further that an overly low frequency operation can result in significant Faradaic losses, also resulting in low $\Delta c_{avg}$.

A corollary to the discussion above is that, for a given CDI cell and flowrate, there exist two frequencies $(f_{low,\Delta c} \textbf{ and } f_{high,\Delta c})$ for which $\Delta c_{avg}$ in a cycle is the same (see Fig. 3d, for example). $f_{high,\Delta c}$ results in less than optimal $\Delta c_{avg}$ because part of the water desalinated in the charging was "re-salinated" prior to efficient extraction of the liquid in the cell (i.e. poor flow efficiency). $f_{low,\Delta c}$ operation efficiently extracts processed water from the cell, but then overly dilutes the effluent fresh water (brine) with feedwater during charging (discharging). Hence, we can interpret operation at the resonant frequency $f_{res}$ (when $f_{low,\Delta c} = f_{high,\Delta c}$) as the optimal tradeoff (to achieve maximum $\Delta c_{avg}$) between these two effects—an operation implying a good balance between properly extracting desalted water versus overly diluting the effluent with feed water. In Appendix A1, we discuss the variation and limits of desalination performance metrics, and practical implications for CDI operation at the resonant frequency.

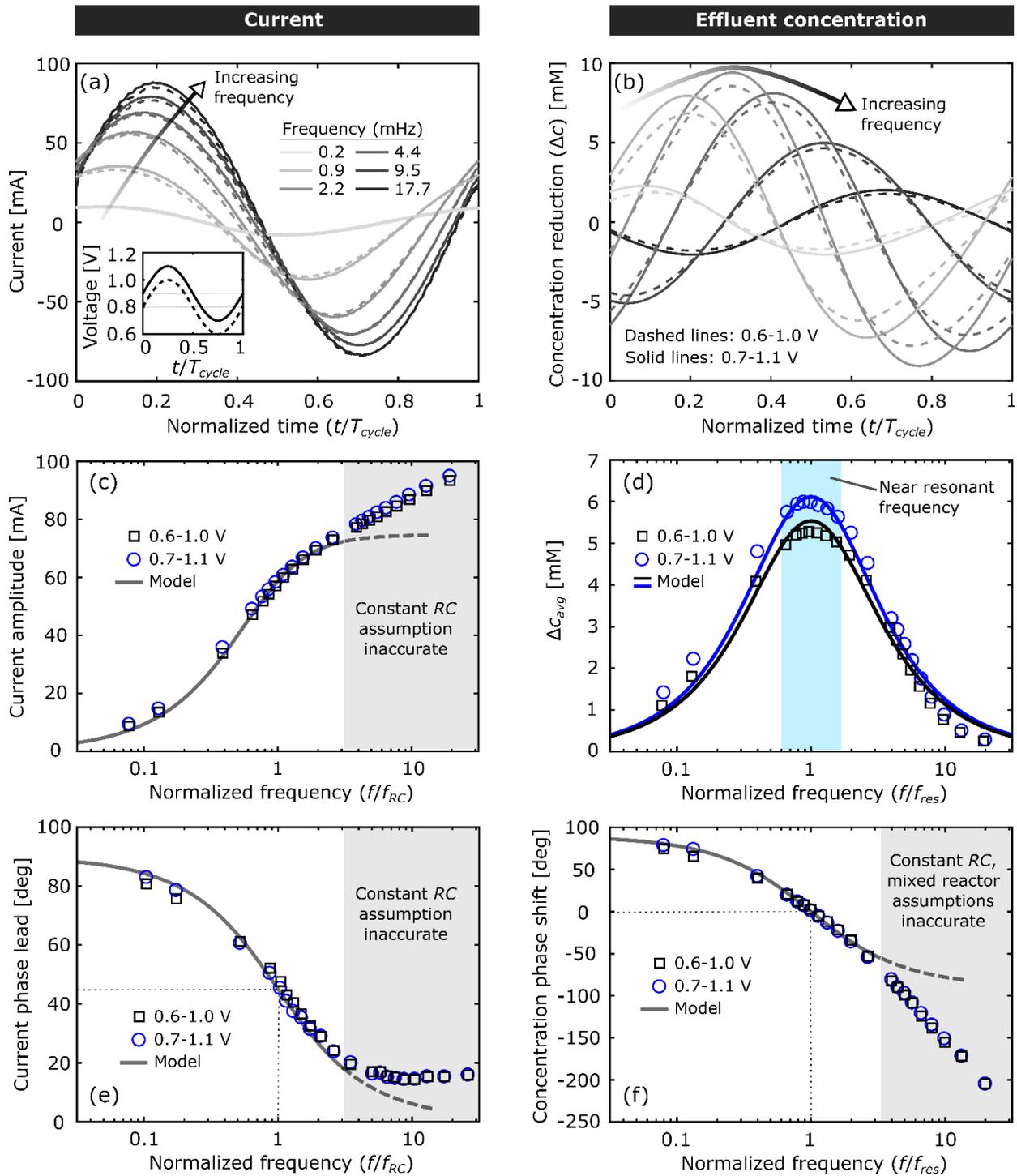

**Figure 3.** Measured (a) current and (b) effluent concentration reduction versus time (normalized by cycle duration) of the CDI cell for a sinusoidal voltage input with frequencies between 0.2 to 17.7 mHz. Solid and dashed lines correspond to operation between 0.6 to 1.0 V ($V_{dc} = 0.8$ V, $\Delta V = 0.2$ V) and 0.7 to 1.1 V ($V_{dc} = 0.9$ V, $\Delta V = 0.2$ V), respectively. The inset in (a) shows the sinusoidal voltage forcing functions. The arrows in (a) and (b) indicate trend of shift of magnitude with increasing forcing frequency. Measured (c) current amplitude, (d) phase of current, (e) average effluent concentration reduction $\Delta c_{avg}$, and (f) phase of effluent concentration versus forcing frequency (same conditions as (a) and (b)). We normalize frequency with $RC$ ($f_{RC}$) and resonant ($f_{res}$) time scales for current and concentration response, respectively. For (c)-(f), solid lines represent model predictions and symbols are experimental data. Operation at resonant frequency results in the maximum $\Delta c_{avg}$ for a given voltage window. Current leads the forcing voltage function at all operating frequencies, and the phase lead is ~45 degrees at the $RC$ frequency $f_{RC}$. The effluent concentration lags (leads) the forcing voltage function for frequencies greater (lower) than the resonant frequency $f_{res}$. At resonant frequency, the effluent concentration is *exactly in phase* with the forcing sinusoid voltage function.

### 4.3 Energy consumption and charge efficiency depend strongly on operating frequency

#### 4.3.1 Energy consumption

First, we study the frequency dependence of the volumetric energy consumption $E_v$ (assuming 100% electrical energy recovery during discharge) defined as

$$E_v \text{ [kWh/m}^3\text{]} = \frac{\int\limits_{t_{cycle}} IV \, dt}{\int\limits_{t_{cycle}|\Delta c > 0} Q \, dt} \ . \tag{9}$$

Figure 4a shows the experimental volumetric energy consumption $E_v$ for a sinusoidal voltage operation versus frequency of operation for voltage windows of 0.6 to 1.0 V ($V_{dc} = 0.8$ V and $\Delta V = 0.2$ V) and 0.7 to 1.1 V ($V_{dc} = 0.9$ V and $\Delta V = 0.2$ V). $E_v$ monotonically decreases as frequency decreases. For a fixed $\Delta V$, a lower $V_{dc}$ (compare data for $V_{dc}$ = 0.8 V and 0.9 V in Fig. 4a) results in smaller $E_v$, but this comes at a price of lower $\Delta c_{avg}$ (see Fig 3d). Note that $E_v$ is very sensitive to even a single decade change in frequency. For example, for $V_{dc}$ = 0.8 V and $\Delta V$ = 0.2 V, at $f / f_{res}$ of 0.1, $E_v$ is 0.015 kWh/m$^3$ and at $f / f_{res}$ of 10, $E_v$ is 0.15 kWh/m$^3$. Clearly, a careful choice of operating frequency and voltage window is important to ensure good trade-off between energy consumption and desalination depth.

Further, to account for salt removal in addition to the corresponding energy consumption, we show in the inset of Fig. 4a the energy normalized adsorbed salt (ENAS) defined as

$$\text{ENAS } [\mu\text{mol/J}] = \frac{\int\limits_{t_{cycle} | \Delta c > 0} Q \Delta c \ dt}{\int\limits_{t_{cycle}} IV \ dt} \ . \tag{10}$$

ENAS is a measure of salt removed (in moles) per energy consumed (in Joules) per cycle. As frequency decreases, ENAS increases, reaches a maximum and then decreases. Importantly, note that the maximum ENAS occurs at a frequency close to (slightly less than) the resonant frequency $f_{res}$, thus again highlighting the importance of operation near the resonant frequency for good overall CDI performance. We attribute the decrease in ENAS at low frequencies to Faradaic energy losses which can become a significant source of energy loss for long cycles (Hemmatifar et al., 2016).

Lastly, we note that our estimate for the volumetric energy consumption $E_v$ in Equation (9) and Figure 4 assumed 100% energy recovery during electrical discharge. In SI Section S6, we show the corresponding volumetric energy consumption values assuming 0% recovery of electrical energy. With 0% energy recovery, we observe the same trends for both $E_v$ and ENAS with frequency and voltage window, as compared with 100% energy recovery.

### 4.3.2 Charge efficiency

We studied the frequency dependence of the conversion of electrical input charge to ions removed as calculated from the effluent stream. We quantify this conversion by defining the cycle charge efficiency as

$$\Lambda_{cycle} = F \frac{\int\limits_{t_{cycle}|\Delta c > 0} Q\Delta c \; dt}{\int\limits_{t_{cycle}|I > 0} I \; dt} \; . \tag{11}$$

Previous studies (Hawks et al., 2018a; Ramachandran et al., 2018) have shown that the cycle charge efficiency $\Lambda_{cycle}$ can be expressed as a product of three efficiencies as $\Lambda_{cycle} = \lambda_{dl}\lambda_c\lambda_{fl} = \overline{\Lambda}\lambda_{fl}$. Here, $\lambda_{fl}$ is the flow efficiency (measure of how well the desalinated or brine water is recovered at the effluent) which depends on number of cell volumes of feed flowed during charging and discharging.

Fig. 4b shows calculated cycle charge efficiency $\Lambda_{cycle}$ values for the same conditions as in Fig. 4a. As frequency decreases, cycle charge efficiency initially increases, reaches a plateau, and then decreases slightly at very low frequency. Also, a larger $V_{dc}$ (and fixed $\Delta V$) results in a higher cycle charge efficiency. We hypothesize that these trends are primarily a result of the frequency dependence of flow efficiency $\lambda_{fl}$, and only a weak function of $\lambda_{dl}$ or $\lambda_c$. Consider that, for finite duration charging cycles at a given flow rate (e.g., $f / f_{res} \sim 0.5$ or less in Fig. 4b), the calculated Coulombic efficiency $\lambda_c$ is high and nearly constant. For example, we estimated a Coulombic efficiency of 0.92 and 0.88 for $V_{dc}$ of 0.8 V and 0.9 V respectively; see SI Section S4 for detailed description of trends in $\lambda_c$. Further consider that, for a fixed voltage window, the EDL efficiency $\lambda_{dl}$ is approximately constant (Ramachandran et al., 2018). For example, from the data of Fig. 3d, we estimate $\lambda_{dl}$ to be 0.8 and 0.91 for $V_{dc}$ of 0.8 V and 0.9 V respectively.

To support our hypothesis, we developed the following analytical expression for flow efficiency $\lambda_{fl}$ for a sinusoid voltage operation:

$$\lambda_{fl} = \frac{1}{\sqrt{1 + (\omega\tau)^2}} \tag{12}$$

The associated derivation is given in Section S1 of the SI. We compared the predicted flow efficiency versus frequency based on Eq.(12) with the corresponding extracted values for flow efficiency values from experimental data ($\lambda_{fl} = \Lambda_{cycle} / \overline{\Lambda} = \Lambda_{cycle} / (\lambda_{dl}\lambda_c)$; see inset of Fig. 4b). Note first from the inset of Fig. 4b that the extracted flow efficiency values from experiments (for both $V_{dc}$ cases) all collapse onto the same curve. Further, our derived flow efficiency expression (Eq. (12)) for sinusoidal voltage operation (dashed line in the inset of Fig. 4b) accurately captures the observed variation in data. This agreement is consistent with an accurate estimate of the mixed reactor cell volume (which is used to evaluate residence time $\tau$ in Eq. (12)).

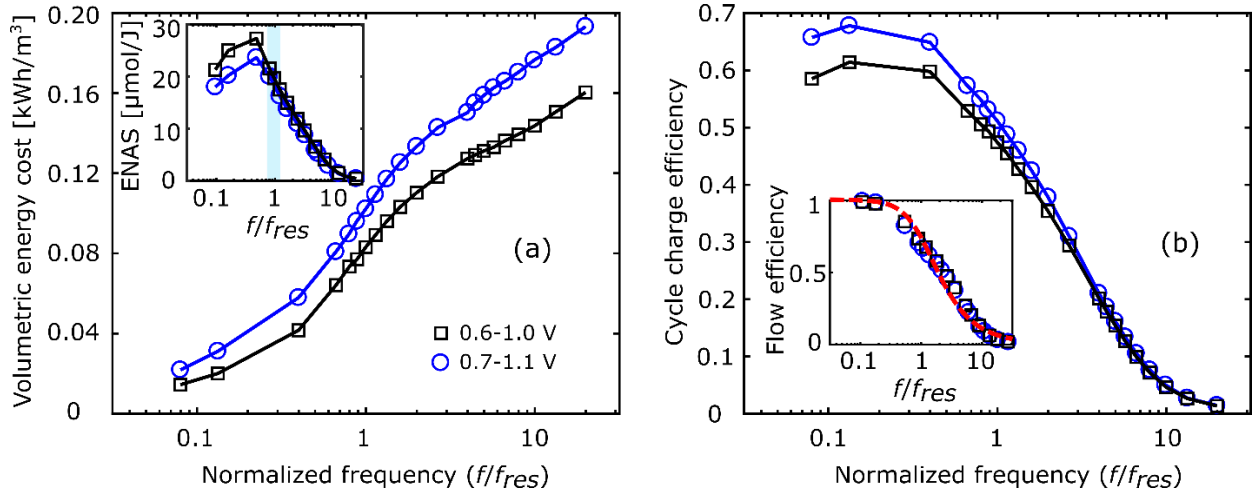

**Figure 4**. Measurements of (a) volumetric energy consumption ($E_v$), and (b) cycle charge efficiency as a function of input sinusoidal voltage frequency for voltage windows of 0.6 to 1.0 V and 0.7 to 1.1 V. Inset of Figure 4a shows the variation of energy normalized adsorbed salt (ENAS) versus input voltage frequency normalized by $f_{res}$. Inset of Figure 4b shows the variation of flow efficiency versus input frequency normalized by $f_{res}$. Dashed line shows model prediction and symbols represent extracted experimental data. $E_v$ and cycle charge efficiency values are higher for the higher voltage window across all frequencies. High frequencies consume the most energy and can result in the least magnitude of effluent concentration reduction (see Fig. 3d).

## 4.4 Generalization of resonant frequency operation for other conventional operations (square and triangular voltage waveforms)

We here generalize the resonant frequency operation for other conventional forcing waveforms such as square voltage (typically referred to as constant voltage operation in CDI) and triangular voltage (an operation similar to constant current operation). We operated the CDI cell with square and triangular voltage waveforms at varying cycle frequencies between 0.7 to 1.1 V (see inset of Figure 5a) and at a constant flowrate of 2.3 ml/min. We used this data to study the variation of performance metrics with applied frequency and waveform shape (see SI Section S5 for current and effluent concentration responses versus time). We then compare the performance of these two voltage waveforms with the sinusoidal voltage waveform at equivalent operating conditions. Figures 5a, 5b, and the inset of 5b show the frequency dependent variation of average concentration reduction $\Delta c_{avg}$, volumetric energy consumption $E_v$, and ENAS, respectively, for square, triangular, and sinusoidal voltage forcing functions to the CDI cell. For both ENAS and $E_v$, we here assume 100% energy recovery during discharge. We refer the reader to SI Section S6 for data corresponding to no energy recovery. As discussed earlier, the upper bound of the voltage window in CDI operation is typically used to avoid significant Faradaic reaction losses, while the lower bound can be used to maintain sufficiently high EDL efficiency. Hence, we here chose to impose the same voltage window (0.7 to 1.1 V) to all three waveforms.

The data of Fig. 5a shows that the square, triangular, and sinusoidal voltage forcing waveforms result in the same general trend for $\Delta c_{avg}$ as a function of frequency. As frequency increases, $\Delta c_{avg}$ initially increases, reaches a maximum, and then decreases at high frequency. All three operating waveforms result in peak values of $\Delta c_{avg}$ near the resonant frequency (indicated by the band of frequencies near $f / f_{res} \approx 1$ in Fig. 5a), highlighting the importance of operation near the resonant time scale.

Of the three waveforms considered here, the square voltage waveform (CV) results in the highest $\Delta c_{avg}$, followed by sinusoidal (less than square wave by ~15%), and then triangular (less than square wave by ~43%) voltage waveforms. However, the volumetric energy consumption $E_v$ for the triangular voltage wave operation is the lowest, followed by sinusoidal (around 1.5x of the triangular waveform $E_v$), and then square (around 4x of the triangular waveform $E_v$) voltage waveforms (see Fig. 5b). The inset of Fig. 5b shows measured ENAS values (a measure of salt removal per energy consumed) for the three waveforms. ENAS values are nearly the same for the triangular and sinusoidal, and their ENAS values are roughly 2x better than that of the square waveform near the resonant operation. We further show in SI Section S6 that for 0% energy recovery during discharge and near resonant operation, ENAS values are highest for sinusoidal waveform, followed by triangular (around 90% of sinusoidal waveform ENAS) and square (around 80% of sinusoidal waveform ENAS) voltage waveforms, respectively.

Together, the data of Fig. 5 and our earlier analysis of sinusoidal operation suggest two important aspects of operational frequency and waveform. First, operation near the resonant time scale (frequency) for these three voltage waveforms yields near optimal values of $\Delta c_{avg}$. Second, the sinusoidal waveform achieves high ENAS (comparable to the triangle voltage waveform), as well as $\Delta c_{avg}$ values much higher than the triangular waveform. Although we have here considered only these three waveforms, we hypothesize these insights span a wide range of both voltage and current forcing function waveforms in CDI. In the next section, we further support this hypothesis using a Fourier mode decomposition of the forcing waveforms.

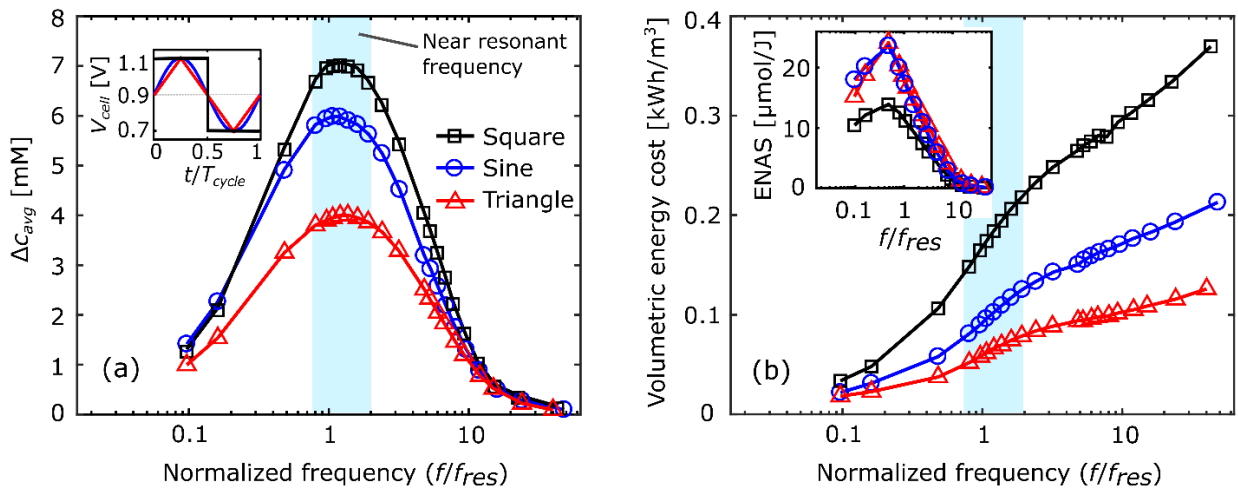

**Figure 5**. Measured values of (a) average effluent concentration reduction, and (b) volumetric energy consumption ($E_v$) as a function of applied voltage frequency normalized by $f_{res}$. Data are shown for three different waveforms: square wave, triangular, and sinusoidal voltages. Wave forms with 0.7 to 1.1 V voltage window are shown in the inset. Inset of Figure 5b shows the variation of energy normalized adsorbed salt (ENAS) versus frequency for the three operations. All operations show a maximum $\Delta c_{avg}$ near the resonant frequency. At resonant operation, the square wave results in maximum $\Delta c_{avg}$, followed by sinusoidal and then triangular voltage operations, but triangular wave consumes the least energy (followed by sinusoidal and then square waves). Figure 5b inset shows ENAS for sinusoidal and triangular voltages are nearly equal and ~2 times higher than square wave operation.

### 4.5 Constructing effluent response for arbitrary forcing functions

We here summarize a Fourier analysis which we find useful in rationalizing the various merits of CDI control schemes. Without loss of generality, we will assume that periodic forcing of the CDI cell is controlled by voltage, although a similar approach can be developed for a current forcing. Eq. (5) in Section 2 is the expression for the effluent response for a sinusoidal forcing voltage with frequency $\omega$ ($= 2\pi f = 2\pi / T$). Any arbitrary voltage forcing $V(t)$ which is periodic with time period $T$ (and phase of zero at $t = 0$) can be decomposed into its Fourier series as

$$V(t) = \frac{a_0}{2} + \sum_{n=1}^{\infty} \left[ a_n \cos\left(n\omega t\right) + b_n \sin\left(n\omega t\right) \right] \tag{13}$$

with Fourier coefficients $a_n$ and $b_n$ given by

$$a_n = \frac{2}{T} \int_0^T V(t) \cos\left(n\omega t\right) dt \quad \textbf{for } n = 0, 1, 2, \dots \ , \tag{14}$$

and

$$b_n = \frac{2}{T}\int_0^T V(t)\sin\left(n\omega t\right)dt \quad \textbf{for } n = 1, 2, \dots \quad .$$ (15)

Each of the term in the summation in Eq. (13) corresponds to a Fourier mode. As shown in Sections 2 and 4.1, CDI can be modeled accurately as a linear time invariant system (under appropriate operating conditions), thus obeying linear superposition of effluent responses due to multiple forcing functions. We thus here hypothesize that the generalized forced response for an arbitrary forcing function in Eq. (13) can be obtained using linear superimposition of responses of its Fourier components (modes). Section 2 presented the frequency response of CDI for a single sine wave and we can now interpret that response as the response of any one of an arbitrary number of Fourier modes.

We here analyze two special cases of Eqs. (13)-(15) corresponding to square and triangular voltage forcing waveforms (as shown in the inset of Fig. 5a). The well-known Fourier modal decompositions for the square ($V_{sq}(t)$) and triangular ($V_{tri}(t)$) voltage waveforms are given by

$$V_{sq}(t) = V_{dc} + \frac{4\Delta V}{\pi}\sum_{n=1}^{\infty}\frac{\sin\left((2n-1)\omega t\right)}{2n-1}$$ (16)

and,

$$V_{tri}(t) = V_{dc} + \frac{8\Delta V}{\pi^2}\sum_{n=1}^{\infty}\frac{(-1)^{(n-1)}\sin\left((2n-1)\omega t\right)}{(2n-1)^2} \quad .$$ (17)

Note that for the triangular wave Fourier modes in Eq. (17), the amplitudes of harmonics decay as $1/(2n-1)^2$, compared to the $1/(2n-1)$ decay for the square waveform (Eq. (16)).

Figure 6 shows the measured effluent concentration response for the square (Figs. 6(a)-(c)) and triangular (Figs. 6(d)-(f)) voltage forcing for flowrate of 2.3 ml/min and an operating frequency spanning 0.43 to 4.3 mHz. In addition, we overlay the effluent response obtained by linearly superimposing the effluent response due to the first two and ten non-zero Fourier modes (excluding the DC component, i.e. up to $n = 2$, and $n = 10$, respectively) in Eqs. (16) and (17). For fair comparison with experiments, we used cycle averaged EDL ($\lambda_{dl} = 0.91$) and Coulombic efficiencies ($\lambda_c = 0.91$ for cases (a), (b), (e) and (f), and $\lambda_c = 0.8$ for (c) and (f)), as per the experimental data (as discussed in Section 4.3).

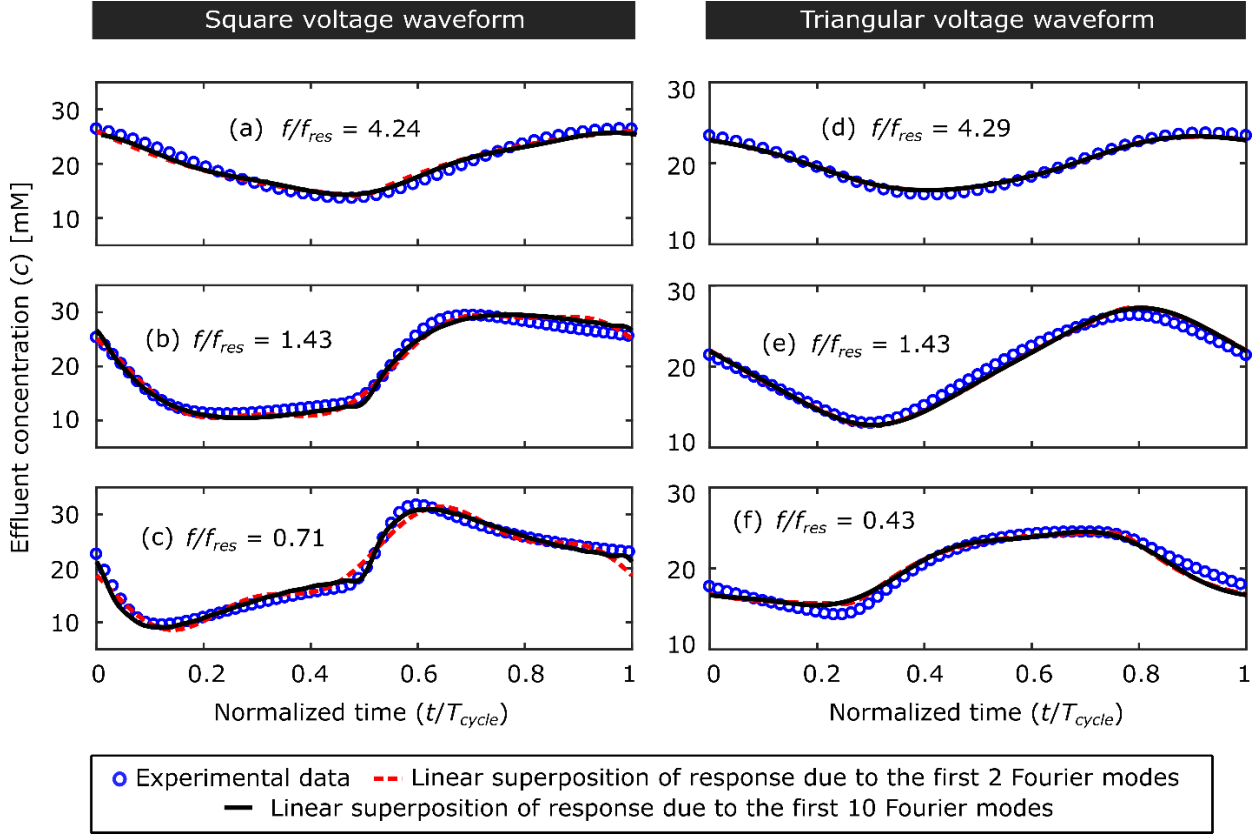

**Figure 6**. Measured effluent concentration versus time normalized by cycle duration for square ((a)-(c)) and triangular ((d)-(f)) voltage forcing at varying frequencies. Data are shown with symbols and the linear superposition of response of the first two and ten Fourier modes based on theory are shown with solid lines. Note that most of the dynamics are well-captured by the first two Fourier modes for the square and triangular waveforms as shown here.

From Fig. 6, we observe that the first two Fourier modes are sufficient to capture the effluent dynamics to a very good approximation for both the square and triangular voltage waveforms and over a practically relevant operating frequency range spanning over a decade. This suggests strongly that the higher harmonics do not contribute significantly to salt removal. In fact, as we will show in Figure 7, inclusion of higher harmonics can sometimes lower $\Delta c_{avg}$ compared to just the fundamental, sinusoidal mode. Briefly, the higher Fourier modes suffer from the drawback of operation at higher fundamental frequency (c.f. Section 4.2.3). Namely, higher modes attempt to force the cell to operate faster than both the RC circuit can respond and faster than water can be recovered from within the cell. Hence, they have inherently inferior flow efficiency and disproportionally consume energy.

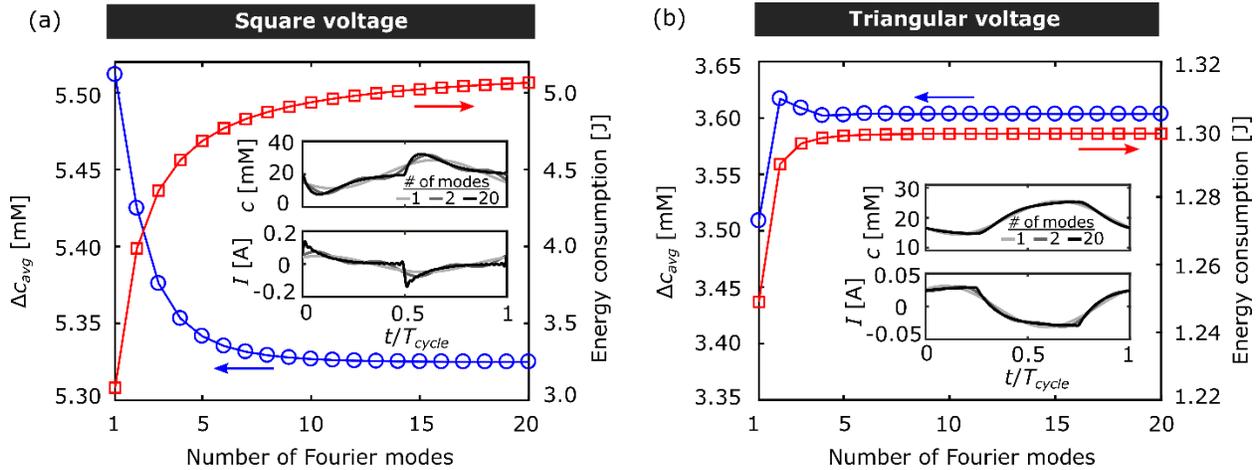

**Figure 7**. Concentration reduction and energy consumption versus the number of Fourier modes used to construct the responses for (a) square voltage, (b) triangular voltage forcing waveforms. Results above are shown for operation at frequency $f = 0.5 f_{res}$, and for values of resistance, capacitance, and flowrate mentioned in Section 3.2. The insets of both plots show effluent

concentration and current responses for the first one, two, and twenty Fourier modes. Effluent concentration dynamics are captured well with just two Fourier modes. Higher modes have negligible effect on effluent concentration waveforms (and average concentration reduction), but consume energy and have inherently poor flow efficiency.

Fig. 7 shows predicted concentration reduction $\Delta c_{avg}$ and energy consumption (estimated here as resistive energy loss in a cycle) versus the number of Fourier modes involved in the summation for the square and triangular waveforms (in Eqs. (16) and (17), respectively). From Figs. 7a and 7b, $\Delta c_{avg}$ does not change significantly beyond the inclusion of the first two to five Fourier modes. This is also apparent in the time variation plots of the effluent concentration presented in the insets of Figure 7. Addition of a second (or higher) Fourier mode can result in either increased $\Delta c_{avg}$ (for e.g., see the triangular voltage case in Fig. 7a) or lower $\Delta c_{avg}$ (for e.g., see the square voltage case in Fig. 7b) compared to the first mode alone depending on the operating frequency. However, the addition of a second (and higher) Fourier mode(s) in the forcing function always results in an increased energy consumption. For example, for the square wave, inclusion of all the modes (here, up to $n = 20$) results in a ~65% increase in energy consumption over the fundamental mode alone. The amplitude of the modes of the triangular waveform decay faster, as $1/(2n-1)^2$, and so their effect on overall energy consumed is less pronounced. For example, including all modes (here, up to $n = 20$) increases energy by only ~5% relative to the fundamental.

For both the square and triangular waves, approximately 95% of the $\Delta c_{avg}$ is achieved by the fundamental (sinusoidal) Fourier mode alone. Adding higher frequency modes therefore provides only a slight increase (or sometimes even a decrease) in salt removal as compared to the fundamental mode alone, but at the great cost of significant energy consumption. This analysis leads us to the hypothesis that, for constant flowrate and appropriately voltage-thresholded operation of CDI, the sinusoidal voltage operation introduced here is likely a near ideal tradeoff between salt removal performance and energy consumption.

## 5. Summary and Conclusions

We developed a model based on a dynamic system approach for CDI. Our analysis considers the coupled effects of electrical circuit response and the salt transport dynamics of a CDI cell. Our study shows that CDI cells with properly designed voltage windows exhibit first-order and near-linear dynamical system response. We performed experiments to validate the model and used both theory and experiments to study CDI performance for a variety of operational regimes. For the first time, we identified an inherent resonant operating frequency for CDI equal to the inverse geometric mean of the $RC$ and flow time scales of the cell. We also quantified the frequency-dependent amplitude and phase of the current and effluent concentration responses for a sinusoidal voltage forcing. We showed using experiments and theory that CDI operation near resonant frequency enables maximum desalination depth $\Delta c_{avg}$.

We further demonstrated that resonant frequency operation can be generalized to other operation methods, and presented analysis of square and triangular voltage forcing waveforms as two relevant case studies. Based on our validated theory, we developed a generalized tool that utilizes Fourier analysis for constructing effluent response for arbitrary input forcing current/voltage waveforms for predicting CDI effluent response. Our work strongly suggests that a sinusoidal forcing voltage for CDI is the ideal operational mode to balance the tradeoff of energy consumption and salt removal in constant flow operation. We believe that our approach can aid in designing and developing methodologies for optimization in CDI performance in the future.

**Acknowledgements**

We gratefully acknowledge funding from the California Energy Commission grant ECP-16-014. Work at LLNL was performed under the auspices of the US DOE by LLNL under Contract DE-AC52-07NA27344. A.R. gratefully acknowledges the support from the Bio-X Bowes Fellowship of Stanford University.

**Appendix A1: Desalination performance at resonant operation**

Encouraged by the strong agreement between experimental observations and model predictions, we here theoretically analyze CDI performance for an applied sine wave voltage that is driven at the cell's resonant frequency (Eq. (7)). Our analyses reveal that energy consumption is minimized and productivity is maximized for a given average concentration reduction when operating at the resonant frequency, thus motivating a more detailed examination of resonant operation desalination performance. To examine these performance relationships, we substitute Eq. (7) into Eqs. (2) and (5), and apply the appropriate integration analysis over a cycle to reveal

$$E_{v,res} = \frac{C\Delta V^2}{\forall}\left(\frac{1}{1+\dfrac{RC}{\tau}}\right) \tag{A1}$$

and,

$$P = \frac{\forall}{2A\tau} \ , \tag{A2}$$

where, $E_{v,res}$ is the volumetric (resistive) energy consumption at resonant operation, $P$ is the throughput productivity (volume of freshwater produced per unit electrode area, per unit time), and the water recovery is implicitly 50% (see Hawks et al., 2018b for a discussion of these metrics).

Note that the expression for volumetric energy consumption (in Eq. (A1)) can be expressed in terms of the harmonic mean of the two time scales $H(\tau, RC) = 2\tau RC/(\tau + RC)$, and this also appears in Eq. (8). (The resonant frequency of operation is still the inverse of the geometric mean of $\tau$ and $RC$.) Moreover, the flow efficiency at resonance $\lambda_{fl,res}$ is obtained as

$$\lambda_{fl,res} = \frac{1}{\sqrt{1 + \dfrac{\tau}{RC}}} \ . \tag{20}$$

Figure A1a and A1b respectively plot several expected performance relationships that follow from Eqs. (A1), (A2), and (8) for operations with fixed voltage window and varying flowrate, and fixed flowrate and varying voltage window. In particular, Figure A1a illustrates the non-linear tradeoff between concentration reduction and productivity for a fixed voltage window and varying flowrate, indicating that a sacrifice in throughput is needed to achieve large concentration reductions (and vice versa) for resonant operation mode. On the other hand, for varying voltage window (and fixed flowrate), higher concentration reduction can be achieved at the expense of energy consumption, as seen in Figure A1b.

Taking the ratio of Eq. (8) to Eq. (A1) yields

$$\frac{\Delta c_{avg,res}}{E_{v,res}} = \frac{2\overline{\Lambda}}{\pi F \forall \Delta V} \ . \tag{21}$$

This relation shows clearly that more efficient operation is achieved for lower voltage windows $\left(\Delta V\right)$ and higher charge efficiencies $\left(\overline{\Lambda}\right)$. Thus, for a fixed concentration reduction, a lower voltage window operation with high capacitance is more efficient than a large voltage window operation with low capacitance. However, due to the finite rate at which the cell can be charged, a larger cell capacitance for a given geometry does not always give a proportionally larger concentration reduction (c.f. Figure A1c). Figure A1c plots concentration reduction and energy consumption as a function of capacitance (or capacity) per CDI reactor fluid volume ($C/\forall$), for a fixed productivity and voltage window. We indicate the design point of the cell used for experiments here using a circular symbol. The figure shows a somewhat surprising result: For a given device geometry (including internal fluid volume) and operating conditions (fixed $\Delta V$ and flowrate), increased capacitance (e.g. due to material improvements) initially improves salt removal performance sharply, but then concentration reduction (and volumetric energy consumption) quickly saturates. The plots of Figure A1 therefore summarize the importance of system level designs for geometry, material, and operational conditions required for a desired CDI performance.

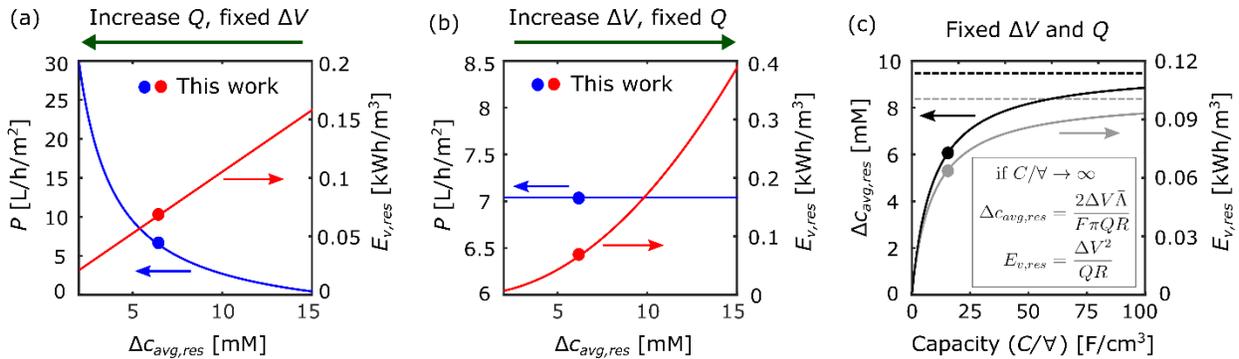

**Figure A1.** The relationships at resonance among energy consumption (Eq. (A1)), concentration reduction (Eq. (8)), and productivity (Eq. (A2)) as a function of (a) flow rate, for fixed voltage window $\Delta V$ of 0.2 V, and (b) voltage window, for fixed flowrate of 2.3 ml/min. (c) Concentration reduction and energy consumption as a function of cell capacitance (or capacity) per CDI reactor fluid volume, for fixed $\Delta V$ of 0.2 V and flowrate of 2.3 ml/min. The dashed lines and inset in (c) show the asymptotic results for concentration reduction and energy consumption when the cell has infinite capacitance/capacity. Data points correspond to the resonant sinusoidal operation presented in Figure 5. We used experimentally determined values of resistance, capacitance, cell volume, and charge efficiency (for $V_{dc}$ of 0.7 V) as mentioned in Sections 3.2 and 4.2.

Supplementary Information for

**Frequency analysis and resonant operation for efficient capacitive deionization**

Ashwin Ramachandran,[a] Steven A. Hawks,[c] Michael Stadermann,[c] Juan G. Santiago [b,*]

[a] Department of Aeronautics & Astronautics, Stanford University, Stanford, California 94305, United States
[b] Department of Mechanical Engineering, Stanford University, Stanford, California 94305, United States
[c] Lawrence Livermore National Laboratory, 7000 East Avenue, Livermore, California 94550, United States

* To whom correspondence should be addressed. Tel. 650-736-1283, Fax 650-723-7657, E-mail: juan.santiago@stanford.edu

# Contents

## S1. Theory for sinusoidal voltage/current forcing for CDI

We here present further details around the theory for predicting desalination dynamics associated with a sinusoidal voltage with a direct current (DC) offset as a forcing for capacitive deionization (CDI) as presented in Section 2 of the main paper. We assume that the electrical response of CDI can be described to a good approximation by a linear series resistor-capacitor (RC) circuit (shown in Figure S1). To describe salt removal and freshwater recovery at the effluent, we assume a continuously stirred tank model. We present our derivation below of the coupled dynamics in two parts. First, we solve for the RC circuit current response for a sinusoidal voltage forcing. Second, we solve for dynamics associated with the effluent concentration reduction using the solution from the previous step, and assuming a well-mixed reactor.

### S1.1 *RC Circuit analysis*

Assume a series RC circuit with a DC-offset sinusoidal forcing voltage given by

$$V(t) = V_{dc} + \Delta V \sin(\omega t),$$  (1)

where $V_{dc}$ is the constant DC component of applied voltage, $\Delta V$ is the amplitude of the sinusoid voltage and $\omega$ is the forcing frequency.

Denoting the capacitive voltage drop by $V_c$, Kirchhoff's voltage law applied to the circuit in Fig. S1 results in

$$RC \frac{dV_c}{dt} + V_c = V(t) = V_{dc} + \Delta V \sin(\omega t) \ .$$  (2)

Equation (2) can be written as,

$$RC \frac{d\tilde{V}_c}{dt} + \tilde{V}_c = \Delta V \sin(\omega t)$$  (3)

where $\tilde{V}_c = V_c - V_{dc}$. For long-duration dynamic steady state operation such that the transient associated with natural response (due to non-zero initial conditions) has decayed, the solution to Equation (3) is described the particular solution. The particular solution to Equation (3) is,

$$\tilde{V}_c(t) = \frac{\Delta V}{\sqrt{1 + (\omega RC)^2}} \sin(\omega t - \arctan(\omega RC))$$  (4)

Since the current given by

$$I = C \frac{dV_c}{dt},$$  (5)

we obtain the current in the circuit from Equation (4) as

$$I(t) = \frac{C\Delta V \omega}{\sqrt{1 + (\omega RC)^2}} \cos(\omega t - \arctan(\omega RC)) = \frac{C\Delta V \omega}{\sqrt{1 + (\omega RC)^2}} \sin\left(\omega t + \frac{\pi}{2} - \arctan(\omega RC)\right). \quad (6)$$

The result in equation (6) can be expressed as

$$I(t) = \Delta I \sin(\omega t + \phi_{IV}) \ , \quad (7)$$

where the current amplitude $\Delta I = \frac{C\Delta V \omega}{\sqrt{1 + (\omega RC)^2}}$ and the phase of current with respect to voltage is given by $\phi_{IV} = \frac{\pi}{2} - \arctan(\omega RC)$.

### S1.2 *Mixed reactor model*

We use a continuously stirred tank reactor model for predicting the effluent concentration dynamics. In a mixed reactor model, the salt removal dynamics is given by

$$\tau \frac{d(\Delta c)}{dt} + \Delta c = \frac{I(t)\overline{\Lambda}}{FQ} \quad (8)$$

where $\tau$ is the flow residence time, and we have assumed constant dynamic charge efficiency, $\overline{\Lambda}$.

Combining Equation (7) in (8), we derive

$$\tau \frac{d(\Delta c)}{dt} + \Delta c = \frac{\Delta I \overline{\Lambda}}{FQ} \sin(\omega t + \phi_{IV}) \ . \quad (9)$$

The solution to Equation (9) is

$$\Delta c(t) = \frac{C\Delta V \omega \overline{\Lambda}}{FQ\sqrt{1 + (\omega \tau)^2}\sqrt{1 + (\omega RC)^2}} \sin\left(\omega t + \frac{\pi}{2} - \arctan(\omega RC) - \arctan(\omega \tau)\right), \quad (10)$$

which can be simplified as

$$\Delta c(t) = \frac{C\Delta V \overline{\Lambda} \omega}{FQ\sqrt{1 + (\omega \tau)^2}\sqrt{1 + (\omega RC)^2}} \sin\left(\omega t + \frac{\pi}{2} - \arctan\left(\frac{\omega(RC + \tau)}{1 - \omega^2 \tau RC}\right)\right) \ . \quad (11)$$

Equivalently,

$$\Delta c(t) = \Delta c_{ac} \sin(\omega t + \phi_{cV}) \ , \quad (12)$$

where $\Delta c_{ac} = \frac{C\Delta V \overline{\Lambda} \omega}{FQ\sqrt{1 + (\omega \tau)^2}\sqrt{1 + (\omega RC)^2}}$ is the maximum change in effluent concentration, and the phase of $\Delta c$ with respect to the forcing voltage $V$ is given by $\phi_{cV} = \dfrac{\pi}{2} - \arctan\left(\omega\tau\right) - \arctan\left(\omega RC\right) = \dfrac{\pi}{2} - \arctan\left(\dfrac{\omega\left(RC+\tau\right)}{1-\omega^2\tau RC}\right)$ . The phase of $\Delta c$ with respect to current is given by $\phi_{cI} = \phi_{cV} - \phi_{IV} = -\arctan\left(\omega\tau\right)$ .

Note further that $\Delta c_{ac}$ and $\Delta I$ are related by,

$$\Delta c_{ac} = \dfrac{\Delta I \bar{\Lambda}}{FQ} \dfrac{1}{\sqrt{1+\left(\omega\tau\right)^2}} \ . \tag{13}$$

## S1.3 *Flow efficiency for sinusoidal forcing*

The number of moles of salt $\Delta N$ removed per cycle is given by

$$\Delta N = \int\limits_{t\,|\,\Delta c > 0} Q\Delta c \ dt = \int\limits_{t\,|\,\Delta c > 0} Q\Delta c_{ac} \sin\left(\omega t + \phi_{cV}\right) \ dt = \dfrac{2}{\omega} Q\Delta c_{ac} \ . \tag{14}$$

In addition, the charge transferred $\Delta q$ to the CDI cell per cycle is given by

$$\Delta q = \int\limits_{t\,|\,I > 0} I \ dt = \int\limits_{t\,|\,\Delta c > 0} \Delta I \sin\left(\omega t + \phi_{IV}\right) \ dt = \dfrac{2}{\omega} \Delta I \ . \tag{15}$$

The cycle charge efficiency $\Lambda_{cycle}$ (measure of moles of salt removed as calculated at the effluent to the electrical charge input in moles) is related to the flow efficiency $\lambda_{fl}$ (measure of fresh water recovery at the effluent) through the following relation,

$$\Lambda_{cycle} = \bar{\Lambda} \ \lambda_{fl} = F\dfrac{\Delta N}{\Delta q} = \dfrac{FQ\Delta c_{ac}}{\Delta I}, \tag{16}$$

where we have used Equations (14) and (15) for the last equality in Equation (16). Substituting Equation (13) in (16), we thus obtain the expression for flow efficiency for the sinusoidal operation as

$$\lambda_{fl} = \dfrac{1}{\sqrt{1+\left(\omega\tau\right)^2}} \ . \tag{17}$$

## S1.4 *Transfer functions for CDI*

We here develop transfer functions relating the output (effluent concentration reduction) to input (current or voltage) for dynamic steady state CDI operation, under appropriate conditions as mentioned in Section 4.1 of the main paper.

Applying a Laplace transform to Equation (2), the transfer function relating the capacitive voltage $V_c$ to the applied voltage $V$ is derived as,

$$\frac{V_c(s)}{V(s)} = \frac{1}{(sRC+1)} \qquad (18)$$

where $s$ is the Laplace variable (Laplace frequency domain).

Further, from Equation (5) we have

$$I(s) = sCV_c(s) \ . \qquad (19)$$

Using Equation (19) in (18), we obtain the transfer function relating the current in the CDI circuit and applied voltage as

$$\frac{I(s)}{V(s)} = \frac{sC}{(RCs+1)} \ . \qquad (20)$$

Next, from the mixed reaction model (Equation (8)), the transfer function relating the effluent concentration reduction to current can be obtained as

$$\frac{\Delta c(s)}{I(s)} = \frac{\overline{\Lambda}}{FQ(\tau s+1)} \ . \qquad (21)$$

Combining Equations (21) and (20), we obtain the following transfer function relating the effluent concentration reduction and the applied voltage:

$$\frac{\Delta c(s)}{V(s)} = \frac{C\overline{\Lambda}}{FQ} \frac{s}{(\tau s+1)(RCs+1)} \ . \qquad (22)$$

Equations (20)-(22) are the transfer functions that relate the input (current or voltage) to the output (effluent concentration reduction) for a linear time invariant CDI system.

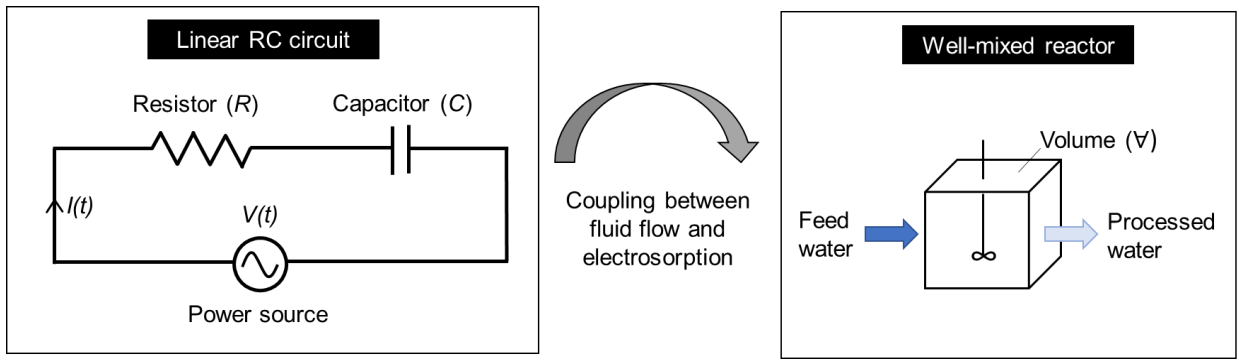

**Figure S1.** Schematic of the model coupling electrical and fluid flow physics in CDI. The linear RC circuit (left) governs ion electrosorption via charge transfer while a well-mixed reactor volume (right) affects the efficiency of recovery of processed water recovery at the effluent via bulk advection.

## S2. Cell resistance and capacitance measurements

We performed a series of preliminary experiments to characterize the CDI cell resistance and capacitance. First, we used simple galvanostatic charging and discharging (see Figure S2) to estimate resistance and capacitance using the following expressions:

$$C_{eq} = \frac{I}{\left( dV / dt \right)} \ ,$$
(23)

and,

$$R_{eq} = \frac{\mid \Delta V \mid_{I \to -I}}{2I} \ ,$$
(24)

where $\mid \Delta V \mid_{I \to -I}$ is the voltage drop when current reverses sign (with the same magnitude). For both cases presented in Fig. S2, using Equations (23) and (24), we estimated a resistance of $2.8 \pm 0.3$ Ohm, and a capacitance of $33.6 \pm 1.7$ F.

To corroborate the cell resistance estimate, we performed electrochemical impedance spectroscopy (EIS) of the entire assembled cell with 20 mM KCl solution and at flow rate of 2.3 ml/min. For EIS measurements (see Figure S3a), we applied a sinusoidal voltage perturbation with amplitude of 10 mV and scanned over a frequency range from 1 MHz to 10 mHz with 0 V DC bias. Using EIS, we estimate an effective resistance of $R_{\infty} \approx 3 \ \text{Ohm}$.

To verify the cell capacitance estimate, we performed cyclic voltammetry for the entire cell. For cyclic voltammetry, we used a scan rate of 0.2 mV/s, flow rate of 2.3 ml/min, and 20 mM KCl solution, and performed measurements till a steady state was reached. In Figure S3b, we show the CV measurement for the fifth cycle (under steady state conditions). Using cyclic voltammetry, we estimate an effective cell capacitance of $C_{eq} \approx 33 \ \text{F}$.

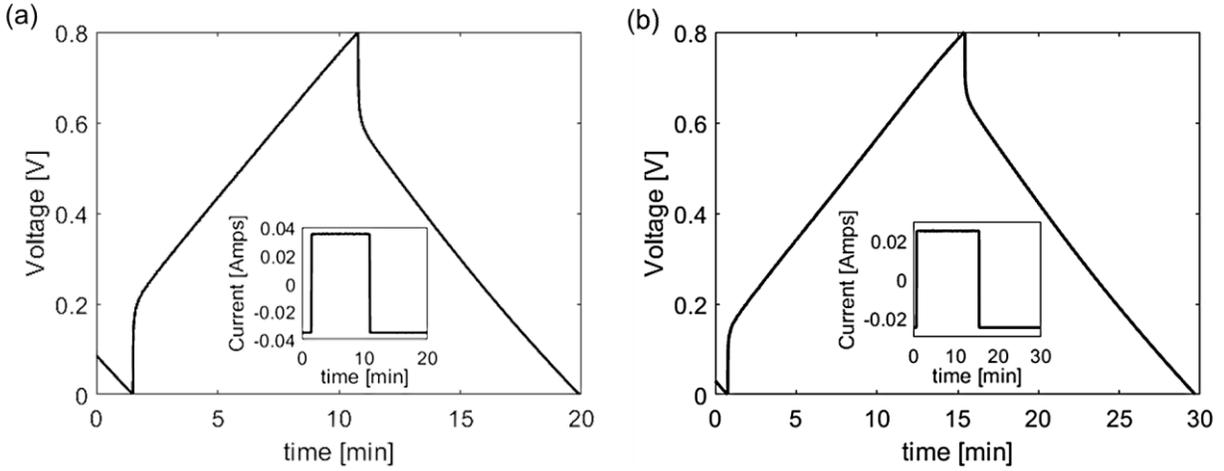

**Figure S2.** Measured galvanostatic charging and discharging (voltage versus time) data for current values of (a) 35 mA, and (b) 25 mA. Insets show the corresponding current versus time data. Data are shown for dynamic steady state operation (fifth charge-discharge cycle), and for a constant flow rate of 2.3 ml/min.

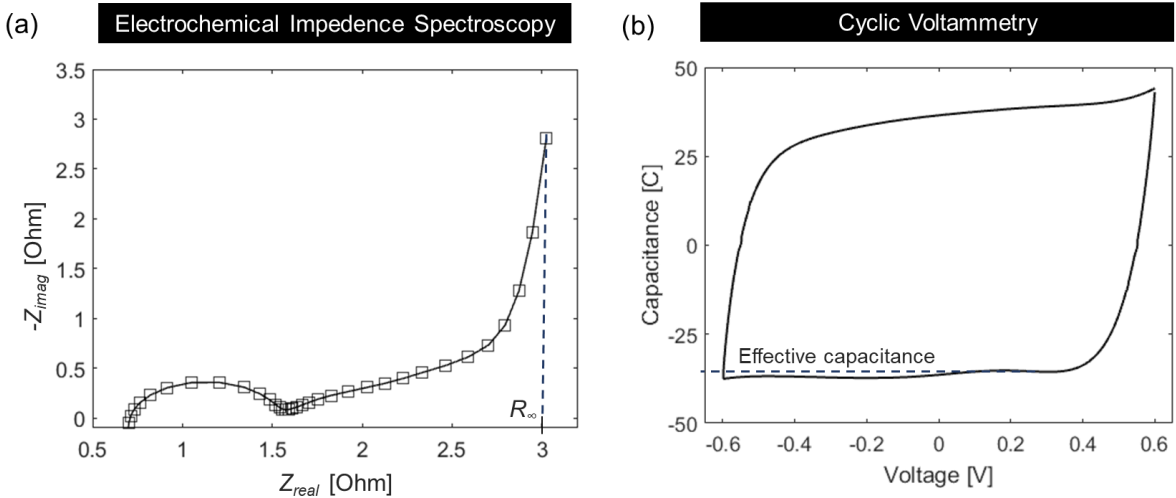

**Figure S3.** (a) Nyquist plot of impedance from electrochemical impedance spectroscopy (EIS) of our cell. We applied a sinusoidal voltage perturbation with amplitude of 10 mV and scanned over a frequency range from 1 MHz to 10 mHz with 0 V DC bias. Highlighted is the estimate of the effective resistance $R_\infty$, which includes the electrode, spacer, contact, and setup resistances. (b) Cyclic voltammogram of our cell performed at a scan rate of 0.2 mV/s, flow rate of 2.3 ml/min and with 20 mM KCl solution. Shown are the data for the fifth cycle (under steady state conditions). We estimate an effective capacitance of $C_{eq} \approx 33\,\mathrm{F}$, and resistance of $R_\infty \approx 3\,\mathrm{Ohm}$.

## S3. Example of an off-design sinusoidal operation

To illustrate an operation wherein the effluent concentration variation with time is not sinusoidal for a DC-offset sinusoidal voltage forcing, we show in Figure S4 a case where the CDI operating voltage varies between 0 to 1.2 V (see Figure S4), at a constant flowrate of 2.3 ml/min. For the results presented in Figure S4, we used a dynamic Gouy-Chapman-Stern (GCS) model which was solved numerically (refer to Biesheuvel et al. (2009) and Ramachandran et al. (2018) for the model and see caption of Fig. S4 for parameters).

Note that the operation presented here allows for the electric double layer (EDL) charge efficiency to vary significantly (see Figure S4d) in a cycle, which violates the constant EDL charge efficiency requirement for a sinusoidal response (see Section 4.1 of the main paper). Thus, the effluent concentration is non-sinusoidal with time (see Fig. S4a). A careful choice of the voltage window (in addition to other conditions as mentioned in Section 4.1 of the main paper) is thus essential to ensure a close-to sinusoidal variation of the effluent concentration with time.

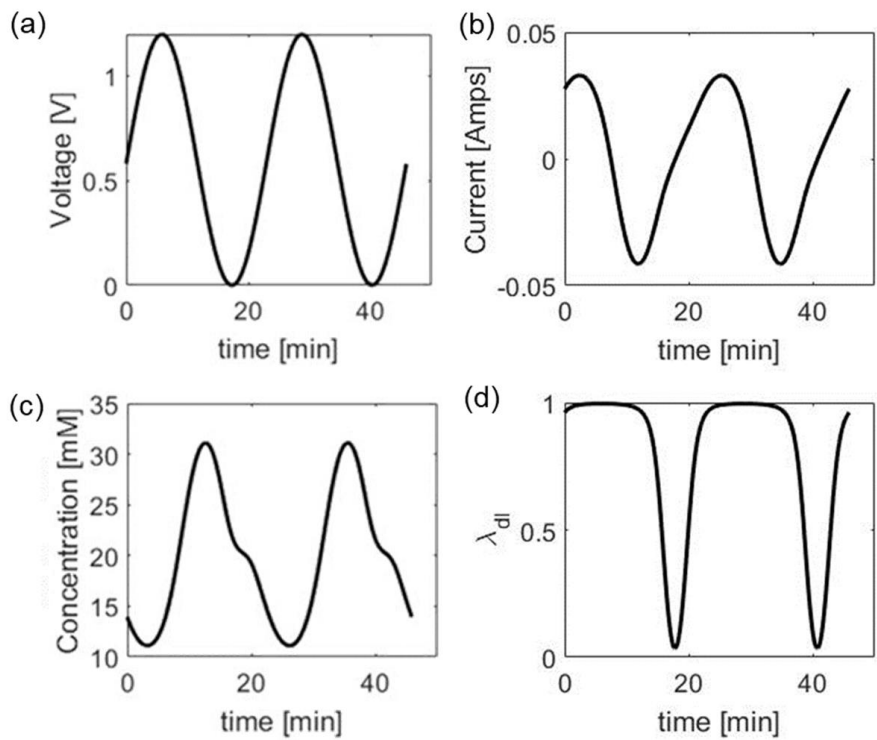

**Figure S4.** Example of an off-design sinusoidal voltage operation. (a) Voltage, (b) current, (c) effluent concentration, and (d) EDL charge efficiency, versus time under dynamic steady state operation (for two consecutive cycles), using a numerical GCS model (see Ramachandran et al., 2018 for more details about the model and notation for parameters). For the GCS model results shown here, we used $c_{st} = 0.4 \text{ F/m}^2$, $a = 100 \text{ m}^2$, $g = 1 \text{ μm/s}$, $R = 1 \text{ Ohm}$, $c_0 = 20 \text{ mM}$, $A = 100 \text{ cm}^2$, $\forall = 2.1 \text{ ml}$, $V_{PZC} = 0 \text{ V}$, and $Q = 2.3 \text{ ml/min}$, with no leakage currents. The operation considered here is a sinusoidal voltage forcing with $V_{dc} = 0.6 \text{ V}$ and $\Delta V = 0.6 \text{ V}$. Note that in this off-design operation, EDL charge efficiency varies significantly during a cycle (between ~ 0 to 1), thus leading to a non-sinusoidal response for the effluent concentration (and current).

**S4. Coulombic efficiency for sinusoidal operation**

We here present the Coulombic efficiency data for sinusoidal voltage operation between 0.6 to 1.0 V, and 0.7 to 1.1 V, supplement to the data presented in Figure 4 of the main paper. Coulombic efficiency $\lambda_c$ is defined as ratio of the recovered electronic charge $q_{out}$ to the input charge transferred $q_{in}$, given by

$$\lambda_c = \frac{q_{out}}{q_{in}} = \frac{\int\limits_{t|I<0} I\ dt}{\int\limits_{t|I>0} I\ dt} \tag{25}$$

Figure S4 shows the Coulombic efficiency versus the forcing sinusoidal voltage frequency for the same operating conditions as presented in Figures 3 and 4 of the main paper. Note that the Coulombic efficiency is relatively constant for moderate to high frequencies (here, greater than around 1mHz), and drops significantly for very low frequencies. Based on the data in Figure S5, we estimated effective Coulombic efficiency values of 0.88 and 0.92 for 0.7-1.1V and 0.6-1.0 V cases, respectively. The drop in Coulombic efficiency at very low frequencies can be attributed to the increased time spent at high voltages during low frequency operations, thus resulting in significant Faradaic charge transfer losses. Also note that the Coulombic efficiency values are lower (i.e. more Coulombic losses) at higher cell voltages.

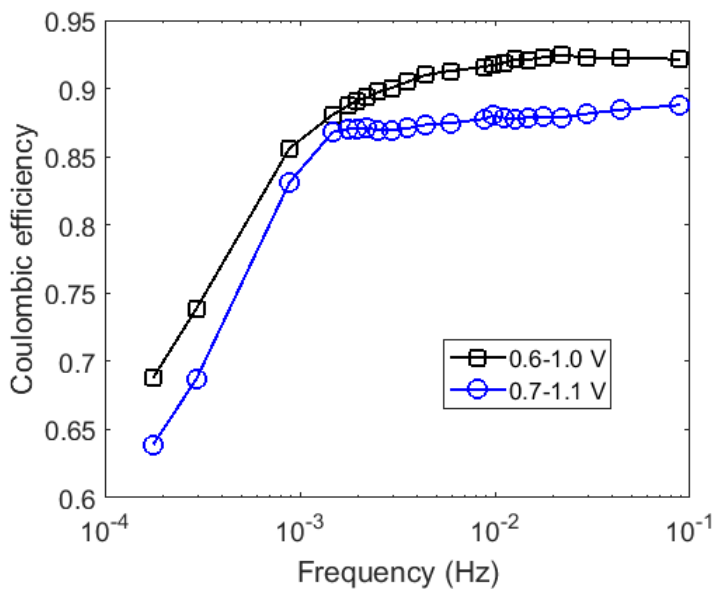

**Figure S5.** Calculated values of Coulombic efficiency $\lambda_c$ as a function of input sinusoidal voltage frequency for voltage windows of 0.6 to 1.0 V and 0.7 to 1.1 V, and constant flowrate of 2.3 ml/min. Coulombic efficiency is nearly constant for moderate to high frequencies, and decreases significantly for low frequencies (long cycle duration). We estimate Coulombic efficiency values (for practical operating frequencies that are not very low) of 0.88 and 0.92 for the 0.7-1.1V and 0.6-1.0 V cases, respectively.

## S5. Measured effluent concentration and current data for square and triangular voltage forcing waveforms at various frequencies

Figure S6 shows measured effluent concentration and current versus time for triangular and square waveform forcing functions corresponding to data presented in Figure 5 of the main manuscript. Here, we show data for a few representative operating frequencies.

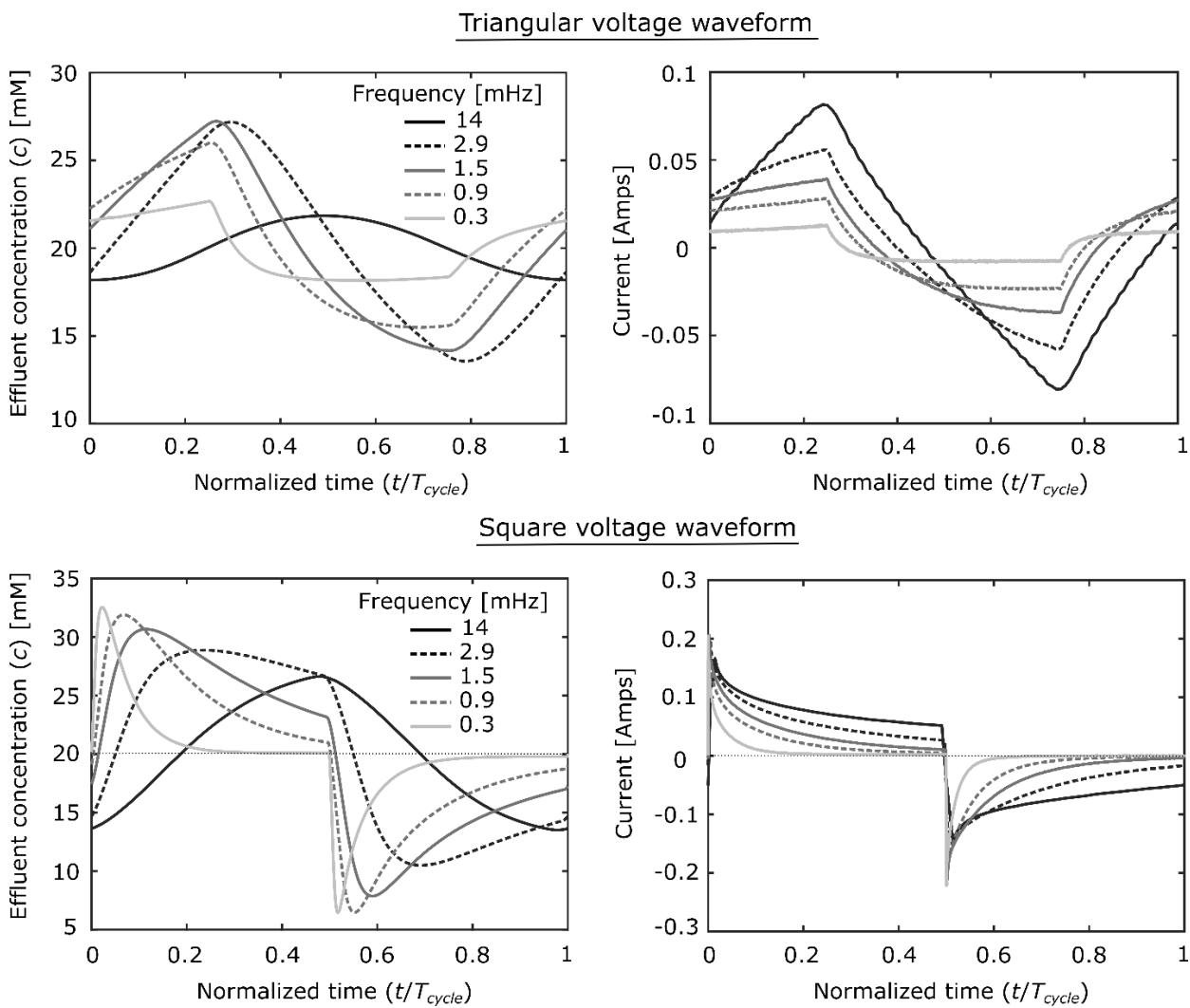

**Figure S6.** Measured values of effluent concentration (left column) and current (right column) for triangular- (top row) and square-wave (bottom row) forcing voltage CDI operation between 0.7 to 1.1 V. Flow rate for all of these experiments was a constant value of 2.3 ml/min.

## S6. Volumetric energy consumption and ENAS with no energy recovery during discharge

We here study the energy consumption metrics (volumetric energy consumption and energy normalized adsorbed salt ENAS) assuming 0% energy recovery during discharge. The volumetric energy consumption with 0% energy recovery $E_v$ and the corresponding energy normalized adsorbed salt ( ENAS) are defined as

$$E_v \ [\text{kWh/m}^3] = \frac{\displaystyle\int_{t_{cycle}|IV>0} IV \ dt}{\displaystyle\int_{t_{cycle}|\Delta c>0} Q \ dt} \ , \tag{26}$$

and

$$\text{ENAS} \ [\mu\text{mol/J}] = \frac{\displaystyle\int_{t_{cycle}|\Delta c>0} Q\Delta c \ dt}{\displaystyle\int_{t_{cycle}|IV>0} IV \ dt} \tag{27}$$

Figure S7 shows the variation of $E_v$ with 0% energy recovery as a function of input sinusoidal voltage frequency for voltage windows of 0.6 to 1.0 V and 0.7 to 1.1 V. $E_v$ decreases with decreasing frequency, whereas ENAS increases, reaches a plateau, and then slightly decreases with decreasing frequency. Further, the 0.6 to 1.0 V voltage window case has lower $E_v$ and higher ENAS values when compared to the 0.7 to 1.1 V case. Note that these trends for $E_v$ are similar to that as observed for volumetric energy consumption $E_v$ and ENAS with 100% energy recovery as presented in Section 4.3.1 of the main paper.

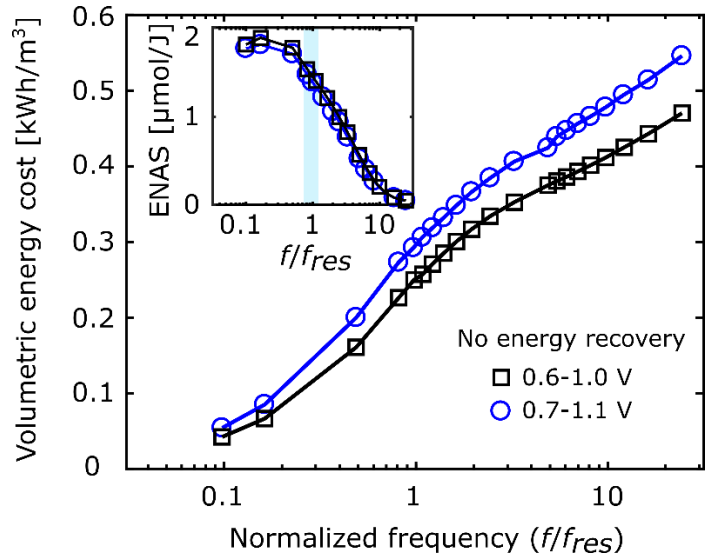

**Figure S7.** Measured volumetric energy consumption with no energy recovery during discharge as a function of input sinusoidal voltage frequency for voltage windows of 0.6 to 1.0 V and 0.7 to 1.1 V. Inset shows the corresponding variation of energy normalized adsorbed salt (ENAS) versus input voltage frequency normalized by $f_{res}$.

In Figure S8, we compare experimental measurements of energy metrics (ENAS and volumetric energy consumption) assuming no energy recovery as a function of operating frequency for three different waveforms: square, triangular, and sinusoidal voltages operated between 0.7 to 1.1 V. We observe from Figure S8 that the volumetric energy consumption with no energy recovery is highest for the square waveform, followed by the sinusoidal, and triangular voltage waveforms respectively. This result is similar to that as seen with 100% energy recovery in Section 4.4 of the main paper.

Further, we see from the experimental data of inset of Fig. S8 that the ENAS values with no energy recovery for frequencies near and lower than the resonant frequency are highest for the sinusoidal waveform, followed by the triangular (less than sinusoidal waveform ENAS by 10%) and square (less than sinusoidal waveform ENAS by 20%) voltage waveforms, respectively.

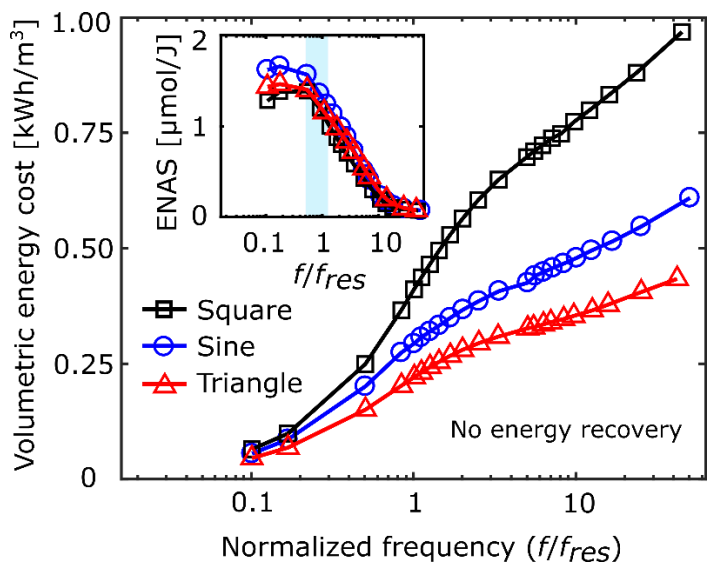

**Figure S8.** Measured values of volumetric energy consumption with no energy recovery during discharge as a function of applied voltage frequency normalized by $f_{res}$. Experimental data are shown for three different waveforms: square wave, triangular, and sinusoidal voltages operated between 0.7 to 1.1 V. Inset shows the corresponding variation of measured energy normalized adsorbed salt (ENAS) versus frequency for the three operations.